\documentclass[%
 twocolumn,
superscriptaddress,prc,
preprintnumbers,
 amsmath,amssymb
]{revtex4-2}
\usepackage{slashed}
\usepackage{cancel}
\usepackage{lipsum}
\usepackage{wasysym}
\usepackage{graphicx}% Include figure files
\usepackage{dcolumn}% Align table columns on decimal point
\usepackage{bm}% bold math
\usepackage[dvipsnames,usenames]{xcolor}
\usepackage{amsmath}
\usepackage{amssymb} % AMS
\usepackage{hyperref}
\usepackage[qm,braket]{qcircuit}
\usepackage{bbold}
\usepackage{algpseudocode}
\usepackage{mathtools}

\newcommand{\veck}{{\bf k}}

\usepackage{subfig}

\begin{document}	

\title{Radiative corrections to 
%$pp$ 
proton-proton fusion in pionless EFT}

\author{Evan Combes}
\affiliation{Department of Physics and Astronomy, University of
  Tennessee, Knoxville, TN 37996, USA} 

\author{Emanuele Mereghetti}
\affiliation{Theoretical Division, Los Alamos National Laboratory, Los Alamos, NM 87545, USA}

\author{Lucas Platter}
\affiliation{Department of Physics and Astronomy, University of
  Tennessee, Knoxville, TN 37996, USA} \affiliation{Physics Division,
  Oak Ridge National Laboratory, Oak Ridge, TN 37831, USA}

\begin{abstract}
We study the leading radiative correction to proton-proton fusion using the pionless effective field theory framework at leading order. We derive the relevant matrix elements and evaluate them using the method of regions. We benchmark the accuracy of our approximations by carrying out numerical computations of the full expressions. We show that the first order radiative corrections due to the exchange of a Coulomb photon between positron and proton-proton systems map onto the Sirlin function and the $\mathcal{O}(\alpha)$ contribution from the Fermi function.
 We furthermore find that the nuclear structure dependent radiative correction omitted in the previous analysis by Kurylov {\it et al} gives an up to 0.2~\% correction to the pp fusion S-factor with its size ultimately depending on a two-nucleon counterterm that renormalizes the axial two-body current $L_{1A}$.
\end{abstract}

\maketitle

\paragraph*{\bf Introduction -}

Proton-proton (pp) fusion is a fundamental process in which two protons combine to form a deuteron, accompanied by the emission of a positron and a neutrino. This reaction initiates the pp chain in the Sun, which is crucial for its energy generation and is responsible for producing the majority of the Sun's neutrinos. The qualitative understanding of this process dates back to Bethe's seminal paper, which described the second branch of the pp chain~\cite{PhysRev.55.434}. Fusion reactions like this are vital for modern stellar simulations~\cite{Acharya:2024lke}. However, experimentally measuring these cross sections is challenging due to the dominant repulsive Coulomb interaction at low energies. Consequently, theoretical calculations with quantified uncertainties are essential for determining these reaction rates. 

Effective field theories (EFTs) have provided a framework for uncertainty estimation in this context. EFTs are low-energy expansions characterized by a parameter, $\epsilon$, which relates to a separation of scales. Two different EFTs have been employed to calculate the pp-fusion rate \footnote{see Refs.~\cite{Epelbaum:2008ga,Hammer:2019poc} and references therein for a detailed discussion of EFTs for the nuclear interaction.}. Chiral EFT uses $\frac{m_\pi, q}{\Lambda_{\chi}}$ as an expansion parameter, where $m_\pi$ is the pion mass, $q$ is the momentum scale inherent to the problem, and $\Lambda_{\chi}$ is a high-energy scale associated with the first relevant degree of freedom not included in this EFT. The expansion parameters for pionless EFT are $ \epsilon =\gamma R$ and $q R$, where $q$ again represents the momentum scale of the process, $\gamma$ denotes the inverse scattering length in the relevant nucleon-nucleon channel, and \(R\) is an estimate for the range of the NN interaction. A naive estimate of the uncertainty is then obtained by assuming that truncating the EFT expansion at order $n$ leads to an uncertainty proportional to $\epsilon^{n+1}$.

Recent chiral and pionless EFT calculations have used the order-by-order convergence pattern to obtain a more reliable uncertainty estimate for the EFT truncation error. For the $pp$ S-factor, these calculations give the values for chiral and pionless EFT of $S(0)=(4.100 \pm 0.024 \pm 0.013 \pm 0.008) \times 10^{-23}$~MeV~fm$^2$~\cite{Acharya:2023xpd} and $S(0)=(4.14\pm 0.01 \pm 0.005\pm 0.006)\times 10^{-23}$~MeV~fm$^2$~\cite{De-Leon:2022omx}, respectively.
The quoted uncertainties combine estimates for the theory error (EFT truncation error), statistical errors introduced in the interaction fitting process, and the uncertainty in the axial coupling $g_A$. 

However, one uncertainty that has not been reliably estimated is the contribution from radiative corrections. Radiative corrections have been previously evaluated in the so-called one-nucleon approximation~\cite{Kurylov:2002vj} that assumes that an additional photon exchange occurs between the positron and the nucleon that also couples to the weak current.
Here, we will compute these one-nucleon corrections but also the {\it nuclear structure} contribution that arises from the photon's coupling to the second nucleon.

Electromagnetic corrections to pp fusion
can be organized in an expansion in $\alpha$ and in the pionless EFT expansion parameter $\epsilon$.
The dominant corrections are Coulomb corrections, that scale as $\alpha m_N/q$ and, at small $q$, need to be resummed into the scattering wavefunction. 
Radiative corrections start contributing at $\mathcal O(\alpha/\pi)$.
Exploiting the fact that, at threshold, the 
electron and neutrino momenta are much smaller than the deuteron binding momentum, we can use the method of regions to separate contributions from photons with different energy and momentum \cite{Beneke:1997zp}.
We will show that:
$(i)$ photons
with low momenta, $|{\bf q}_\gamma| \ll \gamma_t$, where 
$\gamma_t \sim 40$ MeV
is the deuteron binding momentum, induce nuclear-structure-independent corrections that reproduce the Fermi and Sirlin functions \cite{Sirlin:1967zza};
$(ii)$ photons with momentum $|{\bf q}_\gamma| \sim \gamma_t$ induce contributions that depend on nuclear matrix elements-- we will identify corrections that are proportional to the positron energy, scaling as $\mathcal O(\alpha E_e/\gamma_t)$,
and contributions induced by the nucleon magnetic moment and recoil corrections to the one-nucleon vector and axial couplings, which contribute at $\mathcal O(\alpha \epsilon)$; 
$(iii)$ contributions from hard photons, with $|{\bf q}_\gamma| \sim \Lambda \sim R^{-1}$,
are captured by a one-body low-energy constants at $\mathcal O(\alpha/\pi)$ and an electromagnetic correction to $L_{1A}$ at $\mathcal O(\alpha \epsilon)$.
We evaluate the $\mathcal O(\alpha \epsilon)$ corrections in class $(ii)$  and use them to provide an estimate of the uncertainty on the S-factor.

This manuscript is ordered as follows. We will begin by discussing pionless EFT, emphasizing how Coulomb corrections are included. We will then define the S-factor, relate it to the cross-section, and describe the sources of corrections included in its definition. We continue by briefly describing the method of regions, after which we present our results. We end with a brief summary.

\paragraph*{\bf Pionless EFT -}
The pionless EFT Lagrangian at leading order is~\cite{Hammer_2017}:
\begin{eqnarray}\label{eq:Lag0}
    \mathcal L^{{0}}_{\slashed{\pi}EFT} &=& N^{\dagger} \left( i \partial_t
    + \frac{\bm\nabla^2}{2 m_N} \right) N 
     +  t_i^\dagger \Delta_t t_i +  s_a^{\dagger} \Delta_s s_a \nonumber\\
     & &+ y_t \left[t_i^{\dagger} N^T P_i N \right]
     + y_s \left[s_a^{\dagger} N^T P_a N\right] ~,
\end{eqnarray}
where $m_N$ denotes the nucleon mass and $P_{i,a}$ are the spin-isospin projectors for two nucleons in the spin-triplet, isospin-singlet and spin-singlet, isospin-triplet channels, respectively:
\begin{equation}
    P_i = \frac{1}{\sqrt{8}} \sigma_2 \sigma_i \tau_2, \qquad
       P_a = \frac{1}{\sqrt{8}} \sigma_2 \tau_a \tau_2 ~.
\end{equation}
$t_i$ and $s_a$ are the corresponding dibaryon fields; $y_t$, $y_s$ are their non-renormalized couplings to the nucleon fields and $1/\Delta_{t,s}$ their bare propagators. 

\begin{figure}[t]
    \includegraphics[scale=0.9]{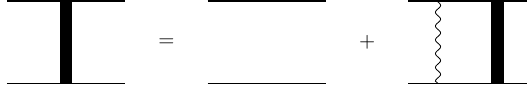}
    \caption{Iterative definition of the Coulomb four-point function. Black boxes represent insertions of $\chi$.}
    \label{fig:CoulombGreenFunction}
\end{figure}

The Coulomb interaction between nucleons arises from minimal substitution and the resulting coupling of the $A_0$ field to the nucleon fields is
\begin{equation}
    \mathcal{L}_{\gamma N} = eN^{\dagger}\frac{1 + \tau_3}{2}N A_{0}~.
\end{equation}
The full Coulomb interaction between incoming protons is included in calculations via insertion of the Coulomb t-matrix, $t_C$, which is related to the Coulomb wavefunction via the relation
\begin{equation}\label{eq:wavefunction}
    \psi^+_{\bf{p}}(\mathbf{k}) = (2\pi)^3\delta^{(3)}(\mathbf{p} - \mathbf{k}) + \frac{t_C(E; \mathbf{k},\mathbf{p})}{E - \frac{\mathbf{k}^2}{2m_N} + i\varepsilon} ~.
\end{equation}
In loops, the protons are moving in the Coulomb fields of one another. This is represented by insertion of the Coulomb four-point function $\chi$ which is defined iteratively in Fig.~\ref{fig:CoulombGreenFunction}~\cite{Ryberg_2016}.
The Coulomb four-point function is related to the Coulomb Green's function $G_C$~\cite{Ryberg_2016}:
\begin{equation}
    \bra{\mathbf{k}}G_C(E)\ket{\mathbf{p}} = -S_{tot}(E,\mathbf{k})\chi(E,\mathbf{k},\mathbf{p})S_{tot}(E,\mathbf{p}) ~,
\end{equation}
where $S_{tot}$ is the two-nucleon propagator,
\begin{equation}
    S_{tot}(E,\mathbf{p}) = \frac{i}{p_0 - \frac{\mathbf{p}^2}{4m_N} + i\varepsilon} ~.
\end{equation}
Additionally, the spectral representation of the Coulomb Green's function is \cite{Kong:2000px}
\begin{equation}\label{eq:greensfunction}
    G_C(E;\mathbf{k}_1,\mathbf{k}_2) = \frac{m_N}{(2\pi)^3} \int d^3 \ell \frac{\psi^*_{\mathbf{\ell}}(\mathbf{k}_1)\psi_{\mathbf{\ell}}(\mathbf{k}_2)}{m_NE - \mathbf{\ell}^2 + i\varepsilon} ~.
\end{equation}
We summarize the discussion of the electromagnetic component to our total Lagrangian:
\begin{eqnarray}
    \mathcal{L}_{EM} &=& N^{\dagger}N [t_C/\chi]\left( \frac{1 + \tau^3}{2} \right)N^{\dagger}N \\
    &&+ eN^{\dagger}\frac12 (1 + \tau_3)NA_{0} + e Q_e \bar{\Psi}_e\gamma^{\mu}\Psi_eA_{\mu} \nonumber ~,
\end{eqnarray}
where $Q_e = -1$ is the electron charge.
The weak interaction is coupled to the leptons via
\begin{align}
    \mathcal{L}_{weak} &= -\sqrt{2} G_F V_{ud} g_V \bar \nu \gamma^\mu \frac{1-\gamma_5}{2} e\,  (J_{\mu}^-)^{1b} ~, \\
    (J_{\mu}^-)^{1b} &= \mathfrak{V}_{\mu}^- - \mathfrak{A}_{\mu}^- ~,
\end{align}
where $\mathfrak{V}_{\mu}^-$ and $\mathfrak{A}_{\mu}^-$ are the leading order isospin lowering weak one-body vector and axial-vector operators, respectively; $G_F = 1.1663788(6) \cdot 10^{-5}$ is the Fermi constant extracted from muon decay \cite{MuLan:2012sih}, $V_{ud}$ is the $ud$ element of the Cabibbo-Kobayashi-Maskawa quark mixing matrix,
and
$g_V$ is the vector coupling constant. At LO in $\alpha$, it is
fixed by current conservation; $g_V = 1$. $g_V$ receives corrections at $\mathcal O(\alpha)$, which are usually subsumed in the ``inner radiative corrections'' $\Delta_R^V$ \cite{Sirlin:1967zza,Sirlin:1977sv,Seng:2018yzq,Seng:2018qru}.
In pionless EFT, the LO vector terms do not contribute due to selection rules and the zeroth axial component is suppressed by $m_N$. Then at leading order we only retain the Gamow-Teller component, $\mathfrak{A}_k^\pm = \frac{g_A}{g_V} N^{\dagger}\tau^{\pm}\sigma_kN$,
with $\tau^{\pm}= (\tau^1 \pm i \tau^2)/2$. 
The axial coupling constant $g_A$ also receives electromagnetic corrections \cite{Gorchtein:2021fce,Hayen:2020cxh,Cirigliano:2022hob}.
Here we will use the experimental determination of $g_A/g_V = 1.2754 \pm 0.0013$ \cite{Workman:2022ynf}, which   
subsumes these corrections. At next-to-leading order, the axial current acquires a two-body counterterm \cite{Kong:2000px}
\begin{equation}
   \mathfrak{A}^\pm_k =  g_A L_{1A} t^{\dagger}_k s^\pm~,
\end{equation}
where $s^\pm = s_1 \pm i s_2$.
As $g_A$, $L_{1A}$ receives electromagnetic corrections from hard photon exchange, we can write
\begin{equation}
    L_{1 A} = L_{1 A}^{(0)} + \alpha L_{1 A}^{(1)} + \ldots~,
\end{equation}
with $L^{(0)}_{1 A} = \mathcal O(\Lambda/m_N)$. We will estimate the size of $L_{1 A}^{(1)}$ by requiring the $pp$ fusion cross section be cut-off independent.
$L_{1 A}$ can be extracted from triton decay \cite{De-Leon:2016wyu}. For such extraction to be used at $\mathcal{O}(\alpha)$, however, the same radiative corrections considered here for pp fusion need to be consistently included in the study of triton decay.

\paragraph*{\bf The S-factor -} 
The capture cross section for charged particles is traditionally expressed in terms of the so-called S-factor. This quantity is obtained by factoring out the Coulomb component from the cross section $\sigma(E)$:
\begin{equation}
    S(E) = E e^{2\pi\eta}\sigma(E)~,
\end{equation}
where $\sigma(E)$ is the cross-section in the center-of-mass frame and $\eta = \frac{m_N \alpha}{2p}$. Ignoring recoil, $\sigma(E)$ for pp-fusion can be written, following the notation for $\beta$ decays,
\begin{align}\label{eq:cs}
    \sigma(E) &=\int \frac{d^3 p_e}{(2\pi)^3 2 E_e} 
    \frac{d^3 p_\nu}{(2\pi)^3 2 E_\nu}
2\pi \delta(E + \Delta m - E_e - E_\nu) \nonumber
\\ & 
 \times \frac{1}{v_{\rm rel}} G_F^2 V_{ud}^2 | \langle d | \mathfrak{A}_- | p p \rangle |^2 \nonumber
\\ & \times F(Z,E_e) (1 + \Delta^V_R)(1 + \delta_R^\prime) (1 + \delta_{\rm NS}) ~,
\end{align}
where $\Delta m = 2m_p - m_d = 0.931$ MeV. The first line includes the kinematics. The second line gives the matrix element in the absence of electromagnetic corrections, except those hidden in the fitted values of $G_F$ and $g_A$,
and those arising from Coulomb interactions between protons.
The third line includes electromagnetic effects, which are usually divided into the Fermi function, $F(Z,E_e)$, which accounts for the distortion of the positron field by the final state deuteron, $\Delta^V_R$, which denotes corrections to the single nucleon vector coupling,
$\delta_R^\prime$, an energy-dependent but nuclear structure independent corrections, and $\delta_{\rm NS}$, which, on the other hand, depends on nuclear structure.
In this paper we will discuss how these corrections arise in the EFT. The discussion is very similar to the EFT treatment of superallowed $\beta$ decays in Refs. \cite{Cirigliano:2024msg,Cirigliano:2024rfk}.

\begin{figure}[t]
\subfloat[\label{fig:kongdiagramA}]{\includegraphics[scale=0.8]{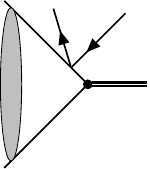}}\hspace{0.3cm}
\subfloat[\label{fig:kongdiagramB}]{\includegraphics[scale=0.8]{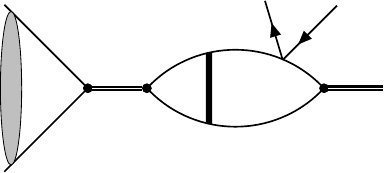}}
\caption{Leading order pp-fusion diagrams with only Coulomb interactions included. Grey ellipses represent insertion of the Coulomb t-matrix.}
\label{fig:kongdiagrams}
\end{figure}

\paragraph*{\bf pp-Fusion at Leading Order}
We start by briefly reviewing the matrix element $\langle d | \mathfrak{A}_- | pp \rangle$.
Kong and Ravndal were the first to study pp-fusion within pionless EFT.
They expressed the leading order amplitude in terms of the Feynman diagrams shown in Fig.\ref{fig:kongdiagrams}. They found the corresponding amplitudes to be, respectively, ~\cite{Kong:2000px} 
\begin{eqnarray}\label{eq:kongeqAB}
    A(p) &=& \sqrt{8\pi\gamma_t}\int \frac{d^3k}{(2\pi)^3}\frac{1}{\veck^2+\gamma_t^2}\psi_{\mathbf{p}}(\veck)~, \\
    B(p) &=& \sqrt{8\pi\gamma_t}\int \frac{d^3k_1}{(2\pi)^3}\frac{d^3k_2}{(2\pi)^3}\frac{G_C(E; \mathbf{k}_1, \mathbf{k}_2)}{\veck_1^2+\gamma_t^2} ~.
\end{eqnarray}
Evaluating the amplitudes in the limit $\mathbf{p} \rightarrow 0$ yields
\begin{eqnarray}
    A(p\rightarrow0) &=& \sqrt{\frac{8\pi C_{\eta}^2}{\gamma_t^3}}e^{\xi}~, \\
    B(p\rightarrow 0) &=& -\frac{m_N^2 \alpha}{4\pi}\sqrt{\frac{8\pi}{\gamma_t^3}}\frac{e^{\xi/2}}{\xi}W_{-1,1/2}(\xi) ~,
\end{eqnarray}
where $\xi = \alpha m_N/\gamma_t$, $W$ is the Whittaker-W function, and $C_{\eta}^2$ is the Sommerfeld factor (see Supplemental Material~\cite{SupplementalMaterials}). The full amplitude, obtained by summing over all rescattering diagrams, was then found to be
\begin{equation}\label{eq:LO}
    T^{\rm LO}_{fi}(p) = g_A \left[ A(p) + \frac{4\pi}{m_N}D_{pp}(p)\psi_{\mathbf{p}}(0)B(p)\right]~,
\end{equation}
where $\psi_{\mathbf{p}}(0)$ is now in coordinate space and where $D_{pp}$ is the proton-proton propagator defined in the Supplemental Materials.

\begin{figure}[t]
\subfloat[\label{fig:diagram1}]{\includegraphics[scale=0.8]{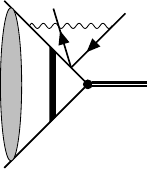}}\hspace{0.3cm}
\subfloat[\label{fig:diagram3}]{\includegraphics[scale=0.8]{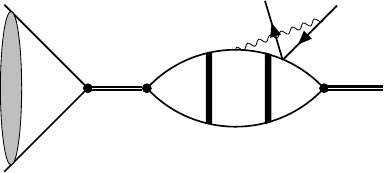}}
\caption{Two diagrams with full Coulomb interactions and single photon exchange. Double thick (thin) lines represent deuteron (proton-proton) dibaryon propagators.}
\label{fig:diagrams13}
\end{figure}

\paragraph*{\bf Electromagnetic corrections to $g_V$ -}
$\Delta_R^V$ includes corrections from 
hard photons, from both the perturbative regime, encompassing the renormalization group evolution from the weak scale, $\mu_{\rm SM} \sim m_W$, to low energy scales of a few GeVs, and nonperturbative contributions from  
the so-called ``$W$-$\gamma$
box''. 
In heavy baryon chiral perturbation theory (HBPT), these corrections are captured by LECs in the one-nucleon $e^2 p$-Lagrangian
\cite{Gasser:2002am,Cirigliano:2022hob,Cirigliano:2023fnz}, which can be re-expressed in terms of current algebra objects \cite{Cirigliano:2023fnz}.
In the EFT \cite{Cirigliano:2023fnz}
\begin{align}
    1 + \Delta_R^V &= \left| g_V(\mu = m_N) \right|^2 \left(1 + \frac{5 \alpha(m_N)}{8\pi}\right) \nonumber  \\ &= 1 + 2.471(25)\%~,
\end{align}
which uses evaluations of the $W$-$\gamma$ box from Refs. \cite{Seng:2018qru,Seng:2018yzq,Czarnecki:2019mwq,Shiells:2020fqp,Hayen:2020cxh,Seng:2020wjq}, as combined in Ref. \cite{Cirigliano:2022yyo}.
$g_V$ is renormalization scale dependent and we have evaluated it at the scale $\mu = m_N$. The renormalization scale dependence is cancelled by the ``ultrasoft'' loops (see discussion on momentum scales). 
In an EFT approach,
it is natural to evaluate $g_V$ at lower scales, $\mu \sim E_e$, which avoids large logarithms in the ultrasoft loops \cite{Cirigliano:2023fnz,Cirigliano:2024msg}. We will discuss the numerical impact of this choice.
All the other corrections in Eq. \eqref{eq:cs} arise from the exchange of dynamical photons in the EFT, as those shown in the diagrams in Fig. \ref{fig:diagrams13}.

\paragraph*{\bf Method of Regions -}
All diagrams at $\mathcal{O}(\alpha)$ can in principle be calculated numerically. However, the method of regions lets us carry out our calculations analytically so that contributions to the Sirlin and Fermi functions can be identified.
The method is based on identifying several distinct scales~\cite{Beneke:1997zp}. In our case, we will separate contributions from photons with different energy and momentum scales. Specifically, we will exploit the scale separation between the external energy $E \sim m_e$ and the deuteron binding momentum $\gamma_t \sim$ 40~MeV to identify three regions:
\begin{itemize}
\item the ultrasoft region, with  $q_0 \sim |{\bf q}| \sim E_e$,
\item the soft region,
with  $q_0 \sim |{\bf q}| \sim \gamma_t$,
\item the potential region, with $q_0 \ll |{\bf q}| \sim \gamma_t$.
\end{itemize}
In practice, we apply the method of regions in the ultrasoft region by dropping the photon momentum with respect to loop momenta that scale as $\gamma_t$; and in the potential region by dropping lepton momentum and mass and photon energy with respect to the photon momentum.
In the ultrasoft regime, 
this approximation induces errors of order $\mathcal O(\alpha E_e^2/\gamma_t^2)$, well below $10^{-5}$.
In the ultrasoft regime, we can show that the virtual emission diagrams factorize the photon momentum dependence out of the hadronic terms. 

\paragraph*{\bf Ultrasoft Radiative Corrections -}
We start by considering contributions from ultrasoft photons. These arise from diagrams such as those shown in Fig. \ref{fig:diagrams13},
in which the integral over the photon energy can be performed without picking nucleon poles, and thus $q^0 \sim |{\bf q}|$. In the ultrasoft regime, 
the momentum scale is set by the electron energy, and thus we can expand the integrand in $|\mathbf q|/\gamma_t$.
The complete set of virtual diagrams in which a photon is exchanged between an electron and a nucleon is presented in the Supplemental Material~\cite{SupplementalMaterials}. In the ultrasoft limit, these diagrams give
\begin{widetext}
\begin{align}\label{eq:usoft}
    T_{\rm us} &=   
   \varepsilon^{* j}  \left\{
     \int \frac{d^d q}{(2\pi)^d}  \frac{ L^j(q)}{q^0 + i \varepsilon}  - i   \int \frac{d^{d-1} q} {(2\pi)^{d-1}} L^j(\mathbf{q})  \right\} T_{fi}^{\rm LO}(p) ~,
\end{align}
\end{widetext}
where $\varepsilon^j$ is the deuteron polarization tensor, $T^{\rm LO}_{fi}$ is given in Eq. \eqref{eq:LO},
and 
\begin{multline}
    L^j(q) = i e^2 \bar u_\nu(p_\nu) \gamma^j P_L (\slashed{p}_e + \slashed{q} - m_e) \gamma^0 v_e(p_e) \\
    \times \frac{1}{\mathbf{q}^2} \frac{1}{(p_e + q)^2 - m_e^2 + i \varepsilon} ~,
\end{multline}
which contains the lepton and photon parts of the diagrams. In the second term in curly brackets in Eq.~\eqref{eq:usoft}, we write $L^j$ as a function of ${\bf q}$ to emphasize that $q^0 = 0$.
In the ultrasoft limit, the photon contribution to the matrix element completely factors out, and the one-photon-exchange radiative correction  yields only a multiplicative factor on the LO matrix element. 
The first term in the second line of Eq. \eqref{eq:usoft} is nothing but the virtual contribution to the Sirlin function in HBPT \cite{Ando:2004rk,Cirigliano:2022hob}.
The second term can be evaluated explicitly and the real part of the diagram yields
\begin{equation}
\delta T_{fi}(p) =    - T^{\rm LO}_{fi}(p) \,\frac{\pi \alpha}{2 \beta}~,
\end{equation}
which, when taking the amplitude squared, corresponds to the $\mathcal O(\alpha)$
term in the Fermi function.
Photon real emission diagrams similarly factorize, so that the contribution of the ultrasoft region reduces to the Sirlin and Fermi functions. We can thus express
\begin{equation}
    \delta_R^\prime = \frac{\alpha}{2\pi} g(E_e,E_0)~, 
\end{equation}
with the Sirlin function
\begin{align}
g(E_e,E_0) &= \frac{3}{2} \log \frac{\mu^2}{m_e^2} - \frac{3}{4} +  \frac{1 + \beta^2}{\beta} \, \log \frac{1+\beta}{1-\beta} \nonumber \\ 
&+4 \left[ \frac{1}{2 \beta} \log \frac{1+\beta}{1-\beta} - 1 \right] \left[ \log \frac{2 \bar E}{m_e}  - \frac{3}{2} + \frac{\bar E}{3 E_e} \right]
\nonumber \\
&+\frac{1}{\beta} \left[ - 4 \,  {\rm Li}_2 \left( \frac{2 \beta}{1 + \beta} \right) - \log^2 \left( \frac{1+\beta}{1-\beta} \right) \right] \nonumber \\
&+ \frac{1}{12 \beta} \left( \frac{\bar E}{E_e}\right)^2  \log \frac{1+\beta}{1-\beta}~,
\end{align}
where $\beta = |\mathbf{p}_e|/E_e$ and $\bar E = E_0 - E_e$. $E_0 = E + \Delta m$ is the electron endpoint energy. Traditionally, the scale $\mu$
is set to $\mu = m_N$, but in the EFT the scale dependence of ultrasoft loops is cancelled by the running of $g_V$, so that we are free to choose any renormalization scale. The residual scale dependence is an indication of the size of missing orders. It can be shown that the same factorization in Eq. 
\eqref{eq:usoft} applies when including NLO interactions, such as the effective range and $L_{1,A}$.  
Corrections to the Sirlin function, arising from diagrams which receive non-negligible potential region contributions, scale as $\mathcal{O}\left( \alpha E_e/\gamma_t \right)$. For each diagram that was predicted to receive potential region contributions, we find numerically that the amplitudes receive about a 2-2.5\% contribution from that region, in line with predictions. Details of these calculations are found in the Supplemental Materials.
Corrections to $\delta_R^\prime$ include $\mathcal O(\alpha^2 \log E_e/\gamma_t )$ terms \cite{Jaus:1970tah,Jaus:1986te,Sirlin:1986cc},
which, in pionless EFT, can be recovered by integrating out soft photon modes and by modifying the renormalization group evolution of $g_V$ 
\cite{Cirigliano:2024msg}. While in $\beta$ decays these can be enhanced by the charge of the nucleus, $Z$, here their scaling is at most $\alpha^2 \sim 10^{-4}$, so that these corrections should be negligible. 
At threshold ($E=0$) and setting the renormalization scale $\mu= m_N$, the correction from the Sirlin function yields \footnote{We use here the ``traditional'' Fermi function, which,
at $\mathcal O(\alpha^2)$,
depends  logarithmically on an arbitrary nuclear radius $R$, set to $R^2 = 5/3 \langle r^2_{\rm ch} \rangle$. The dependence on $R$ can be understood as dependence on a renormalization scale  \cite{Hill:2023acw,Hill:2023bfh,Borah:2024ghn},
which is cancelled by an $\mathcal O(\alpha^2)$ 
nuclear-structure-dependent contribution arising from the integration of soft photon modes \cite{Cirigliano:2024msg}. For $pp$ fusion, these corrections are numerically small, and we neglect them. Our calculation thus accurately captures terms of $\mathcal O(\alpha^2 \pi^2)$, but misses $\mathcal O(\alpha^2 \log R)$ and $\mathcal O(\alpha^2)$ terms. 
}
\begin{equation}
  \frac{ \int d {E_e} E_e p_e (E_0 - E_e)^2 F(Z,E_e) (1 + \delta_R^\prime)}{\int d {E_e} E_e p_e (E_0 - E_e)^2 F(Z,E_e) } =1.0163~, 
\end{equation}
so that the combined ``inner'' and ''outer'' corrections shift the cross section by
\begin{align}
        (1 + \Delta_R^V) (1 + \delta_R^\prime) &\rightarrow 1.04141(25)~,
\end{align}
with the error determined by the nonpertubative input in the $W$-$\gamma$ box \cite{Seng:2018qru}. 
This is in good agreement with what is used, for example, in Ref. \cite{Acharya:2023xpd}.
Setting the renormalization scale to $\mu = 2 E_0$, we obtain
\begin{equation}
    (1 + \Delta_R^V) (1 + \delta_R^\prime) \rightarrow 1.04156(25)~.
\end{equation}
The difference, $1.5 \cdot 10^{-4}$, is an estimate of missing $\mathcal O(\alpha^2)$ contributions.

\paragraph*{\bf Leading Nuclear Structure Correction $\delta_{\rm NS}$ -} 
Nuclear structure dependence arises from photons with momentum $|{\bf q}| \sim \gamma_t$, which can
be either soft or potential.
The soft region does not contribute at the order we are working, but starts contributing at $\mathcal O(\alpha^2)$. The potential region only contributes to diagrams in which the $q^0$ integration is forced to pick poles in the nucleon propagators. These are shown in the Supplemental Materials~\cite{SupplementalMaterials}. Since the photon three-momentum is $\mathcal O(\gamma_t)$, in this region we can expand the diagrams in powers of the lepton energies and masses over $|{\bf q}|$. If we use the leading order weak and electromagnetic currents, the only contributions that do not vanish between $S$-wave states are proportional to the electron energy or the proton relative momentum. These corrections scale as $\mathcal O( \alpha E_e/\gamma_t) \sim 10^{-4}$. Similar terms were constructed in Ref. \cite{Cirigliano:2024msg} for superallowed $\beta$ decays. 
Larger corrections can be obtained by using $\mathcal O(1/m_N)$ corrections to the weak axial- and vector-currents and to the nucleon-photon vertices. These diagrams induce corrections that scale as $\mathcal O(\alpha \gamma_t/m_N)$. We focus here on diagrams that are enhanced by the relatively large nucleon magnetic moments, and neglect other recoil corrections. This will be sufficient to provide an idea of the size of $\delta_{\rm NS}$. 
Corrections from the nucleon magnetic moments and from weak magnetism are shown in the Supplemental Materials. They induce a contribution to the amplitude
of the form
\begin{multline}\label{eq:mag0}
     T_{\rm pot} =  
\, \frac{e^2}{4\pi} \left( 
\left(T^{(0)}_{\rm mag} + T_{\rm ct}\right)\, \delta^{j \ell} +  T^{(2) j \ell}_{\rm mag}  \right) \\
    \times \bar u_\nu(p_\nu) \gamma^j P_L  v_e(p_e) \varepsilon^{*\ell}  ~,
\end{multline}
with
\begin{widetext}
\begin{align}\label{eq:magnetic}
    T^{(0)}_{\rm mag} & =\sqrt{8 \pi \gamma_t}  
    \Bigg\{   
\int \frac{d^{d-1} k_1}{(2\pi)^{d-1}}
 \int   \frac{d^{d-1} k_2}{(2\pi)^{d-1}} \frac{1}{{\bf k}_1^2 + \gamma_t^2} \psi_{\bf p}({\bf k}_2) V^{(0)}_{\rm mag}({\bf k}_1 -{\bf k}_2) \nonumber \\
    & \hspace{0.5cm}+  \frac{4 \pi}{m_N} D_{pp}(p) \psi_{\bf p}(0)  
\int \frac{d^{d-1} k_1}{(2\pi)^{d-1}} \int
\frac{d^{d-1} k_3}{(2\pi)^{d-1}}
 \int   \frac{d^{d-1} k_2}{(2\pi)^{d-1}}
 \frac{1}{{\bf k}_1^2 + \gamma_t^2} V^{(0)}_{\rm mag}({\bf k}_1 -{\bf k}_3) G_C(E; {\bf k}_3, {\bf k}_2)   \Bigg\}~, \\
 T_{\rm ct} &= \sqrt{8 \pi \gamma_t}  \left(- \frac{g_A}{m_N} L_{1 A}^{(1)}  D_{pp}(p) \psi_{\bf p}(0) \right)~, \label{eq:ct}
\end{align} 
\end{widetext}
and an analogous expression for the tensor component $T^{(2) j \ell}$. Note that in Eq.~\ref{eq:ct}, $\psi_{\mathbf{p}}(0)$ is in coordinate space.
After projecting in spin and isospin, the magnetic moment induced potential acting in the $^1S_0$-$^3S_1$ channel is 
\begin{equation}
    \hspace{-0.203cm}V_{\rm mag}^{(0)}({\bf k})  = -\frac{4\pi}{ m_N {\bf k}^{\, 2}} 
    \left(\frac{g_A}{3} (\kappa_0 + \kappa_1)
    + \frac{1}{6} (\kappa_0 - \kappa_1) \right)~,
\end{equation}
with $\kappa_1 = 4.71$ and $\kappa_0 = 0.88$ the nucleon isoscalar and isovector magnetic moments, respectively. 
The tensor component of $V^{(2) j\ell}_{\rm mag}$
is given in the Supplemental Materials, but its matrix element vanishes in the limit of ${\bf p}\rightarrow 0$,
so that the leptonic structure of Eq. \eqref{eq:mag0} becomes identical to the LO, and this correction can be interpreted as a shift in the nuclear matrix element, justifying Eq. \eqref{eq:cs}. Equation \eqref{eq:magnetic} corresponds to the matrix element of $V_{\rm mag}^{(0)}$ between the LO pionless EFT deuteron and $pp$ wavefunctions.
By replacing $G_C(E;{\bf k}_1, {\bf k}_2)$ with the free Green's function 
$G^{(0)}(E;{\bf k}_1, {\bf k}_2)$, it is easy to see that the integral on the second line of \eqref{eq:magnetic}
is UV divergent: 
\begin{align}
   \left. T^{(0)}_{\rm mag}\right|_{\rm uv} &= \sqrt{8 \pi \gamma_t}\, \frac{1}{m_N}  D_{pp}(p)\psi_{\bf p}(0)    \log \frac{\Lambda}{\gamma_t} \nonumber \\
    &  \left(\frac{g_A}{3} (\kappa_0 + \kappa_1)
    + \frac{1}{6} (\kappa_0 - \kappa_1) \right),
\end{align}
which, with Eq. \eqref{eq:ct}, shows that the UV divergence can be absorbed by an electromagnetic shift to $L_{1A}$.
At leading order in pionless EFT, we can thus write 
\begin{equation}
    \delta_{\rm NS} = 2 \alpha \left(T^{(0)}_{\rm mag} + T_{\rm ct} \right) \left[ T^{\rm LO}_{fi}(p) \right]^{-1}~.
\end{equation}
In the numerical evaluations, we will set $L^{(1)}_{1 A}$ to zero, and evaluate $T^{(0)}_{\rm mag}$ for a range of cut off between $m_\pi/2$ and $3 m_\pi/2$. The electromagnetic shift to $L_{1 A}$ could in principle be extracted from $^3$H decay, once the electromagnetic corrections in Eq. \eqref{eq:magnetic} are included 
in the Gamow-Teller matrix element. 

$\delta_{\rm NS}$ can be evaluated numerically in the S-wave, in the limit $\mathbf{p} \rightarrow 0$, and yields a per-mil level correction of approximately $-0.0011$ to $-0.0018$, with the aforementioned cutoffs.
We notice that with LO chiral EFT deuteron wavefunctions, the matrix element in Eq. \eqref{eq:magnetic} would be finite \cite{Beane:2001bc}, and $L_{1 A}^{(1)}$ would only be needed at higher order. It would thus be interesting to repeat the calculation in chiral EFT.

\paragraph*{\bf Summary -} We examined explicitly the single-photon exchange contributions to pp-fusion in pionless EFT. Previous calculations were carried out in the one-nucleon approximation and neglected contributions from photon exchanges between positron and the nucleon that does not couple to the weak current. We utilized the method of regions and found that the dominant ultrasoft photon contributions to the scattering amplitude factorize the photon momentum dependence from the hadronic terms. The photon contributions reproduce the $\mathcal{O}(\alpha)$ piece of the Fermi function and the Sirlin function \cite{Sirlin:1967zza} in HBPT while the hadronic term is given by the LO amplitude originally derived by Kong \& Ravndal. 
We evaluated also, for the first time, the leading nuclear structure contribution arising from radiative corrections explicitly. This contribution arises from $\mathcal{O}(\alpha)$ diagrams that include the exchange photon's magnetic coupling to the nucleon or Coulomb photons paired with subleading weak operators. The sum of these diagrams is UV divergent. However, using a cutoff of order $m_\pi$, we found that $\delta_{\rm NS}$ contributes a correction of at most $0.2\%$ to the pp-fusion S-factor.

We conclude, therefore, that the largest uncertainty in current pp S-factor calculations still arises from the EFT truncation of the nuclear Hamiltonian, uncertainty in the parameters that are fitted to experimental data and the axial coupling constant $g_A$.

\begin{acknowledgments}
We thank V.~Cirigliano, W.~Dekens, J.~de~Vries and M. Hoferichter for insightful conversations and comments on the manuscript.
This work was supported by the
National Science Foundation (Grant No. PHY-2111426), the Office of Nuclear Physics, US Department of Energy (Contract No. DE-AC05-00OR22725), and by
Los Alamos National Laboratory's Laboratory Directed Research and Development program under projects
20210190ER and 20210041DR.
Los Alamos National Laboratory is operated by Triad National Security, LLC,
for the National Nuclear Security Administration of U.S.\ Department of Energy (Contract No.\
89233218CNA000001).
    We acknowledge support from the DOE Topical Collaboration ``Nuclear Theory for New Physics,'' award No.\ DE-SC0023663.
\end{acknowledgments}

\bibliography{bibliography}

\end{document}

\title{Supplemental Materials: Radiative corrections to proton-proton fusion in pionless EFT}

\author{Evan Combes}
\affiliation{Department of Physics and Astronomy, University of
  Tennessee, Knoxville, TN 37996, USA} 

\author{Emanuele Mereghetti}
\affiliation{Theoretical Division, Los Alamos National Laboratory, Los Alamos, NM 87545, USA}

\author{Lucas Platter}
\affiliation{Department of Physics and Astronomy, University of
  Tennessee, Knoxville, TN 37996, USA} \affiliation{Physics Division,
  Oak Ridge National Laboratory, Oak Ridge, TN 37831, USA}

\begin{abstract}
This supplemental material provides a brief discussion on the application of the method of regions for ultrasoft and potential photons. A comparison of the ultrasoft region reduction vs. full expression, evaluated numerically, is given for one diagram in the S-wave channel. All diagrams that were evaluated in the ultrasoft region are presented.
\end{abstract}

\maketitle
\section{LO Amplitude}
Kong \& Ravndal use a contact interaction formulation of the pionless EFT Lagrangian:

\begin{eqnarray}
    \mathcal{L} &= N^{\dagger}\left( \partial_t + \frac{\bf{\nabla}^2}{2m_N} \right)N + C_0(N^T P_s N)(N^T P_s N)^{\dagger}~,
\end{eqnarray}
where $P_s$ is the same as the ${}^1S_0$ projector defined in the main text. The LO pp-fusion amplitude is expressed
\begin{equation}
    T_{fi}(p) = g_A\left[ A(p) + \frac{\psi_{\mathbf{p}}(0)}{C_0^{-1} - J_0(p)}B(p) \right] ~,
\end{equation}
where the individual amplitudes, the corresponding diagrams of which are displayed in the main text, are respectively,
\begin{align}
    A(p) &= \sqrt{8\pi \gamma_t}\int \frac{d^3k}{(2\pi)^3}\frac{\psi_{\mathbf p}(\mathbf k)}{\mathbf k^2 + \gamma_t^2}~,\\
    B(p) &= \sqrt{8\pi \gamma_t}\int \frac{d^3k}{(2\pi)^3}\frac{d^3k'}{(2\pi)^3}\frac{G_C(E; \mathbf k, \mathbf k')}{\mathbf k^2 + \gamma_t^2} ~,
\end{align}
with $G_C$ the Coulomb Green's function,
\begin{equation}\label{eq:GC}
    G_C(E; \mathbf k , \mathbf k') = m_N \int \frac{d^3q}{(2\pi)^3}\frac{\psi_{\mathbf q}(\mathbf k)\psi^*_{\mathbf q}(\mathbf k')}{\mathbf p^2 - \mathbf q^2 + i\epsilon} ~,
\end{equation}
and $\frac{\psi_{\mathbf{p}}(0)}{C_0^{-1} - J_0(p)}$ arises from summing over all possible rescattering diagrams. $|\psi_{\mathbf{p}}(0)|^2$ is the Sommerfeld factor,
\begin{equation}
    C_{\eta}^2 = \frac{2\pi \eta}{e^{2\pi\eta}-1}~,
\end{equation}
where $\eta = \frac{\alpha m_N}{2p}$. Writing the amplitude in full,
\begin{multline}
    T_{fi}(p) = g_A\sqrt{8\pi \gamma_t} \left\{ \int \frac{d^3k}{(2\pi)^3}\frac{\psi_{\mathbf p}(\mathbf k)}{\mathbf k^2 + \gamma_t^2}  \right. \\
    + \left. \frac{\psi_{p}(0)}{C_0^{-1} - J_0(p)}\int \frac{d^3k}{(2\pi)^3}\frac{d^3k'}{(2\pi)^3}\frac{G_C(E; \mathbf k, \mathbf k')}{\mathbf k^2 + \gamma_t^2} \right\} ~.
\end{multline}
Replacing the bare couplings with renormalized ones, 

\begin{align}
    \frac{1}{C_0^{-1}(\mu) - J_0(p)} \nonumber &=  \frac{4 \pi}{m_N} D_{pp}(p) \\
    &= \frac{4\pi}{m_N}\left[ \frac{1}{a_p}+\alpha m_N H(\eta)  \right]^{-1} ~,
\end{align}
where $a_p$ is the proton-proton scattering length, measured to be $a_p = -7.82$ fm, and 
\begin{equation}
    H(\eta) = \psi(i\eta) + \frac{1}{2i\eta} - \ln(i\eta) ~.
\end{equation}
As $H(\eta)$ vanishes in the limit $\mathbf{p} \rightarrow 0$, we use

\begin{equation}
     \frac{1}{C_0^{-1}(\mu) - J_0(p)} = \frac{4\pi}{m_N} a_p ~.
\end{equation}

\section{Ultrasoft Region Calculations}

\begin{figure}[t]
    \subfloat[\label{fig:diagram1}]{\includegraphics[clip,width=0.2\columnwidth]{Diagram1}}\hspace{0.5cm}
    \subfloat[\label{fig:diagram2}]{\includegraphics[clip,width=0.2\columnwidth]{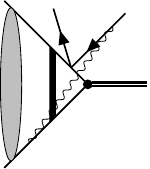}}

    \subfloat[\label{fig:diagram3}]{\includegraphics[clip,width=0.45\columnwidth]{Diagram3}}\hspace{0.2cm}
    \subfloat[\label{fig:diagram4}]{\includegraphics[clip,width=0.45\columnwidth]{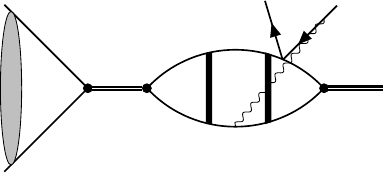}}

    \subfloat[\label{fig:diagram5}]{\includegraphics[clip,width=0.45\columnwidth]{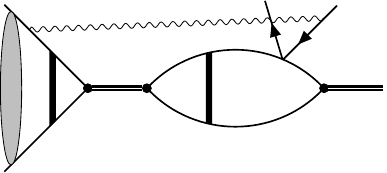}}\hspace{0.2cm}
    \subfloat[\label{fig:diagram6}]{\includegraphics[clip,width=0.45\columnwidth]{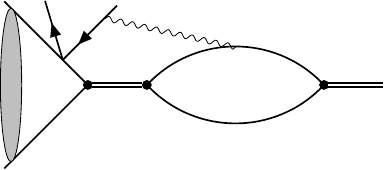}}

    \subfloat[\label{fig:diagram7}]{\includegraphics[clip,width=0.862742\columnwidth]{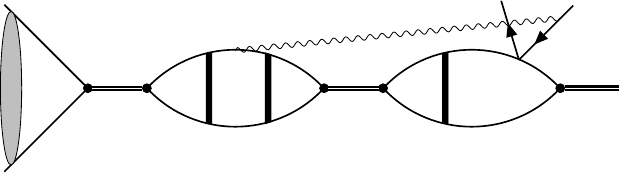}}

    \subfloat[\label{fig:diagram8}]{\includegraphics[clip,width=0.862742\columnwidth]{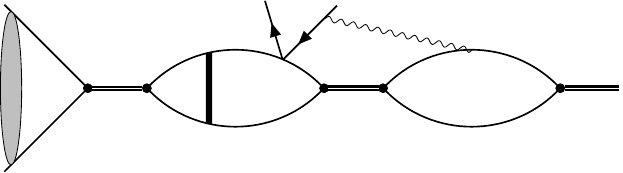}}
    \caption{All single-photon exchange diagrams considered in this work. Grey ellipses represent insertion of the Coulomb t-matrix; black boxes represent insertions of the Coulomb four-point function $\chi$; and double thick (thin) lines represent deuteron (proton-proton) dibaryon propagators.}
    \label{fig:alldiagrams}
\end{figure}

All calculations performed in this work follow a similar pattern. Let us examine the amplitude corresponding to Fig.\ref{fig:diagram1}. We drop overall constants here and highlight just the procedure. With $k_1^0$ and $k_2^0$ already integrated over, we obtain the amplitude,

\begin{multline}\label{eq:amplitude1}
	i\mathcal{A}_1 \propto \int \frac{d^4q}{(2\pi)^4}\int \frac{d^3k_2}{(2\pi)^3} \int \frac{d^3k_1}{(2\pi)^3} L^j(q)\frac{\psi_{\mathbf{p}}(\veck_1)}{E' - \frac{\veck_2^2}{m_N} + i\varepsilon} \\
    \times G_C\left(E + q^0 - \frac{\vecq^2}{4m_N};\veck_1 - \frac{\vecq}{2}, \veck_2 - \frac{\vecq}{2} \right) ~. 
\end{multline}
We then have to separate the ultrasoft from the potential region contributions. In this diagram, we are not forced to pick poles in the nucleon propagators, so we expect to be able to exclude the potential region. 

The ultrasoft region is defined as the region where $q_0 \sim |{\bf{q}}| \sim E_e$, and $q_0$ and $\bf q$ are thereby considered to be small in comparison to the other scales appearing in the loop integral. In practice, we therefore drop $\bf{q}$ with respect to $\bf{k}_1$ and ${\bf k}_2$. The energy $\mathbf{q}^2/m_N$ is very small and is also dropped. On the other hand, $m_N q^0 \sim \gamma_t^2$. We use the spectral representation of the Coulomb Green's function in Eq.~\ref{eq:GC} to reduce Eq.~\ref{eq:amplitude1} to
\begin{multline}
    i\mathcal{A}_1(\text{us}) \propto \int \frac{d^4q}{(2\pi)^4}\int \frac{d^3k_2}{(2\pi)^3}  L^j(q) \frac{(2\pi)^3m_N }{E' - \frac{\mathbf{k}_2^2}{m_N} + i\epsilon}\\
	\times \int d^3\ell \frac{\psi_{\mathbf{\ell}}(\mathbf{k_2})}{m_N(E + q^0) - \mathbf{\ell}^2 + i\epsilon} \\
    \times \int \frac{d^3k_1}{(2\pi)^3}\psi^*_{\mathbf{\ell}}(\mathbf{k_1})\psi_{\mathbf{p}}(\mathbf k_1) ~.
\end{multline}
This expression, together with the completeness relation of the Coulomb wavefunctions, reduces in the limit $\mathbf{p} \rightarrow 0$ to
\begin{equation}
    i\mathcal{A}_1(\text{us}) \propto \int \frac{d^4q}{(2\pi)^4}\int \frac{d^3k_2}{(2\pi)^3} \frac{L^j(q)}{q^0 + i\varepsilon}\frac{\psi_{\mathbf{p}}(\veck_2)}{\gamma_t^2  + \veck_2^2 -  i\varepsilon} ~.
\end{equation}
In the above equation, we leave $\frac{1}{q^0 + i\varepsilon}$ as is because there is no other scale present in that term. The integral over $q$ is the single nucleon correction and the integral over $k_2$ is the same as obtained by Kong \& Ravndal. Thus, we can construct an analytic solution using the results presented in Ref.~\cite{Kong:2000px}.

While Eq. \ref{eq:amplitude1} is difficult to solve analytically, we can solve it numerically. We proceed by Fourier transforming from momentum to coordinate space. We find

\begin{multline}
    \hspace{-0.1cm}i\mathcal{A}_1 \propto \int \frac{d^4q}{(2\pi)^4}\int d^3r_1 \int d^3r_2L^j(q) \psi_{\bf{p}}(\vecr_1)e^{-i \frac{\vecq}{2} \cdot\vecr_1}e^{i \frac{\vecq}{2} \cdot\vecr_2} \\
    \times \frac{e^{-\gamma_t r_2}}{4\pi r_2} G_C\left(E + q^0 - \frac{\vecq^2}{4m_N}; -\vecr_2, \vecr_1 \right) ~. 
\end{multline}
We will only consider the S-wave amplitude. Then we take only the S-partial wave term in the complex exponentials. We shift the coordinate integrals to spherical polar form and are left with

\begin{multline}
    i\mathcal{A}_1^{S} \propto \int \frac{d^4q}{(2\pi)^4}\int dr_1 \int dr_2 L^j(q) \psi_{p}(r_1) \\
    \times \frac{\sin(r_1 q/2)}{r_2 q/2}\frac{\sin(r_2 q/2)}{r_2q/2}\frac{e^{-\gamma_t r_2}}{4\pi r_2} \\
    \times G_C^{(0)}\left(E + q^0 - \frac{\vecq^2}{4m_N}; r_2, r_1 \right)~. 
\end{multline}
Note that $\psi_p(r)$ now indicates the S-wave Coulomb wavefunction. This can be further simplified via contour integration over $q^0$.
In addition, we use the bound-state S-wave Coulomb Green's function in terms of regular and irregular Coulomb functions,~\cite{Ryberg_2016}
\begin{multline}
    G_C^{S}(-B; r', r) = -\frac{m_Np}{4\pi}F_0(\eta, \rho_<) \\
    \times \frac{\left[ iF_0(\eta,\rho_>) + G_0(\eta, \rho_>) \right]}{\rho_<\rho_>} ~,
\end{multline}
where $\rho = pr$ and the subscripts denote the lesser or greater of $\rho$ and $\rho'$. So shifting the photon momentum integral also to spherical polar form, we obtain
\begin{multline}
    i\mathcal{A}_1^{S} \propto \int dq\int dx \int dr_1 dr_2 \frac{\slashed{p}_e + \gamma^0(q^0)' - \gamma^3qx + m_e}{\sqrt{E_e^2 + q^2 + 2qp_ex} + i\epsilon} \\
	\times \psi_{p} (r_1)\frac{\sin(r_2q/2)}{q}e^{-\gamma_t r_2} \frac{\sin(r_1q/2)}{q} \frac{r_1}{p' r_< r_>} \\
    \times F_0(\eta',p'r_{<})\left[ iF_0(\eta',p'r_>) + G_0(\eta', p'r_>) \right]~, 
\end{multline}
where $x$ is the cosine of the angle between $\vecq$ and $\bf{p_e}$, and 
\begin{align}
p' &= \sqrt{m_N}\sqrt{E + q^0 - \frac{\vecq^2}{4m_N}} ~, \\\eta' &= \frac{\alpha m_N}{2p'} ~, \\
(q^0)' &= -E_e - \sqrt{E_e^2 + q^2 + 2q p_e x} + i\varepsilon ~.
\end{align}
$(q^0)'$ is the pole chosen in the $q^0$ contour integration. The above integral can now be solved numerically and
we find that the terms with no $q$-dependence in their numerators in the ultrasoft and full expressions differ by 1.2-1.4\% when the positron energy is maximized. 

\begin{figure}[t]
    \includegraphics[scale=0.65]{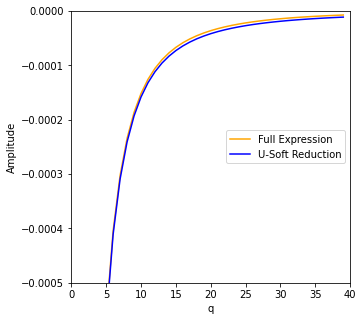}
    \caption{Numerical results for terms with no $q$-dependence in their numerators in the complete and ultrasoft region amplitudes of the diagram in Fig.\ref{fig:diagram1}. Overall constants dropped and evaluation performed at the maximum allowed positron energy.}
\end{figure}

\begin{figure*}[t]
\centering 
\subfloat[\label{fig:magdiagram2-1}]{\includegraphics[clip,width=0.3\columnwidth]{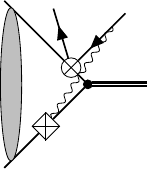}}\hspace{1cm}
\subfloat[\label{fig:magdiagram2-2}]{\includegraphics[clip,width=0.3\columnwidth]{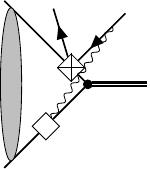}}\hspace{1cm}
\subfloat[\label{fig:magdiagram2-3}]{\includegraphics[clip,width=0.3\columnwidth]{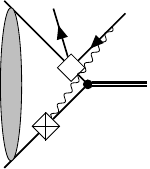}}

\subfloat[\label{fig:magdiagram4-1}]{\includegraphics[clip,width=0.65\columnwidth]{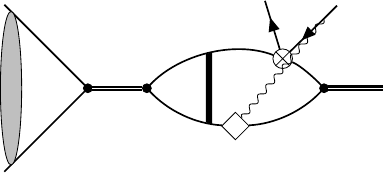}} \hspace{0.4cm}
\subfloat[\label{fig:magdiagram4-2}]{\includegraphics[clip,width=0.65\columnwidth]{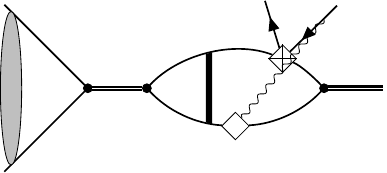}} \hspace{0.4cm}
\subfloat[\label{fig:magdiagram4-3}]{\includegraphics[clip,width=0.65\columnwidth]{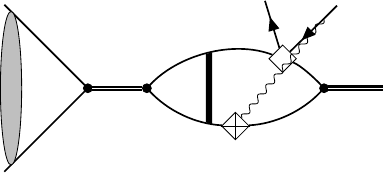}}
\caption{All diagrams considered for magnetic corrections to pp-fusion. Circles (diamonds) with crosses represent the spatial component of the axial-vector (vector) weak operators; empty diamonds represent the time component of the vector weak operators.}
\label{fig:magneticdiagrams}
\end{figure*}

\section{\label{potentialregion}Potential Region Contributions}
\iffalse
\begin{figure}[t]
\subfloat[\label{fig:diagram2}]{\includegraphics[scale=0.8]{Diagram2}}
\subfloat[\label{fig:diagram4}]{\includegraphics[scale=0.8]{Diagram4}}
\caption{The two diagrams which receive soft region contributions.}
\label{fig:diagrams24}
\end{figure} \fi
Two diagrams, seen in Figs.~\ref{fig:diagram2} and \ref{fig:diagram4}, are predicted to receive potential region contributions of order $\mathcal{O}\left( \frac{\alpha E_e}{\gamma_t} \right)$. This was determined by expanding the full expression for each and subtracting off the potential region expressions while dropping terms that are much smaller than $\mathcal{O}\left( \frac{\alpha E_e}{\gamma_t} \right)$. 

Potential region contributions arise from the Coulomb Green's function. The expansion of the Coulomb Green's function should have a one-to-one matching onto an arbitrary number of photon exchange diagrams. However, this is found not to be the case. Specifically, for Figs.~\ref{fig:diagram2} and \ref{fig:diagram4}, or more generally for diagrams in which photon couplings are to the nucleons which do not undergo weak decay, this matching fails for zero photon exchange. This is a consequence of the derivation of the Coulomb Green's function, which entails a specific choice of poles. Subsequently, for those diagrams, there are additional potential region contributions. The iterative definition of the Coulomb Green's function, up to next-to leading order, is
\begin{multline}
    G_C(E;\mathbf{p}_1,\mathbf{p}_2) = (2\pi)^3\delta^{(3)}(\bf{p}_1 - \bf{p}_2) \\
    + \frac{e^2(\mathbf{p_1} - \mathbf{p_2})^{-2}}{E - \frac{\bf{p_1}^2}{m_N} +
    i\varepsilon}\frac{1}{E - \frac{\bf{p}_2^2}{m_N} + i\varepsilon}~,
\end{multline}
where $e$ is the elementary charge. Comparing this against one- and zero-photon exchange in all diagrams in Fig.~\ref{fig:alldiagrams}, it has been verified that only the diagrams Figs.~\ref{fig:diagram2} and \ref{fig:diagram4} receive additional potential region contributions.

The amplitude corresponding to Fig.~\ref{fig:diagram2} is, with overall constants dropped and without the additional term,
\begin{eqnarray}
    i\mathcal{A}_2 =&& \int \frac{d^4q}{(2\pi)^4}\int \frac{d^3k_2}{(2\pi)^3} \int \frac{d^3k_1}{(2\pi)^3} L^j(q)  \psi_{\bf{p}}(\veck_1) \\
    && \times \frac{G_C\left( E + q^0 + \frac{\bf{q}^2}{4m_N}; \veck_2 - \frac{\bf{q}}{2}, \veck_1 - \frac{\bf{q}}{2} \right)}{E' - \frac{1}{m_N}(\veck_2 - \mathbf{q})^2 + i\varepsilon} \nonumber ~,
\end{eqnarray}
which would naively lead us to assume smaller potential region contributions. The additional contribution is calculated to be, with overall constants similarly dropped,
\begin{eqnarray}
    i\delta A_2 &=& m_N^2\int \frac{d^4q}{(2\pi)^4}\int \frac{d^3k_1}{(2\pi)^3}L^j(q)\frac{\psi_{\mathbf{p}}(\veck_1)}{\gamma_t^2 + (\veck_1 - \mathbf{q})^2 - i\varepsilon} \nonumber \\
    && \times\frac{1}{\gamma_t^2 + m_Nq^0 + \left( \veck_1 - \frac{\mathbf{q}}{2} \right)^2 + \frac{\mathbf{q}^2}{4} - i\varepsilon} ~.
\end{eqnarray}
Numerical results reveal soft contributions of 2-2.5\% for both diagrams.

\section{Magnetic Contributions}
Magnetic contributions to the pp-fusion process arise from one-body leading order weak vector and axial-vector operators,~\cite{Butler_2001}

\begin{align}
	\mathfrak{V}_0 &= N^{\dagger}\frac{1 + \tau^a}{2}N~, \\
	\mathfrak{A}_0 &= i g_A N^{\dagger} \tau_a \frac{\mathbf{\sigma}\cdot (\overleftarrow{\nabla} - \overrightarrow{\nabla})}{m_N} N~, \\
	\mathfrak{V}_k &= iN^{\dagger} \frac{1 + \tau_a}{2} \frac{(\overleftarrow{\nabla}_k - \overrightarrow{\nabla}_k)}{2m_N}N \label{eq:magnvector} \\
    &\hspace{0.5cm} - N^{\dagger}(\kappa^0 + \kappa^1\tau_a)\epsilon_{kij}\frac{\sigma_i (\overleftarrow{\nabla}_j + \overrightarrow{\nabla}_j)}{4m_N}N ~,\nonumber \\
	\mathfrak{A}_k &= g_AN^{\dagger}\frac{\tau^a}{2}\sigma_kN ~,
\end{align}
where the isospin matrix $\tau^a$ is $\tau^3$ at vertices involving photon coupling to the nucleon and $\tau^-$ for weak vertices. Note that in Eq.~\ref{eq:magnvector}, there is a factor of $1/4$ in the second term. This is a result of our choice of convention for the nucleon isoscalar and isovector magnetic moments. The combinations of operators used are those that contribute at $1/m_N$. They are shown in Fig.~\ref{fig:magneticdiagrams}.

These diagrams yield the leading nuclear structure corrections. In addition, they induce a tensor component, which vanishes if the proton relative momentum goes to zero, ${\bf p} \rightarrow 0$. For completeness, we provide the expression of the tensor amplitude and potential here:
\begin{widetext}
\begin{align}\label{eq:magnetic2}
    \mathcal A^{(2) j \ell}_{\rm mag} & =\sqrt{8 \pi \gamma_t} \, \frac{e^2}{4\pi} 
    \Bigg\{   
\int \frac{d^{d-1} k_1}{(2\pi)^{d-1}}
 \int   \frac{d^{d-1} k_2}{(2\pi)^{d-1}} \frac{1}{{\bf k}_1^2 + \gamma_t^2} \psi_{\bf p}({\bf k}_2) V^{(2) j \ell}_{\rm mag}({\bf k}_1 -{\bf k}_2) \nonumber \\
    & + \frac{4 \pi}{m_N} D_{pp}(p)  \psi_p(0)  
\int \frac{d^{d-1} k_1}{(2\pi)^{d-1}} \int
\frac{d^{d-1} k_3}{(2\pi)^{d-1}}
 \int   \frac{d^{d-1} k_2}{(2\pi)^{d-1}}
 \frac{1}{{\bf k}_1^2 + \gamma_t^2} V^{(2) j \ell}_{\rm mag}({\bf k}_1 -{\bf k}_3) G_C(E; {\bf k}_3, {\bf k}_2)   \Bigg\}, \\
     V_{\rm mag}^{(2) j \ell}({\bf k}) &= -\frac{4\pi}{ m_N {\bf k}^{\, 2}} \left(  \frac{k^j k^\ell}{{\bf k}^{\, 2}} - \frac{\delta^{j \ell}}{3} \right)   \frac{1}{4}\left(g_A (\kappa_0 + \kappa_1) - (\kappa_0 - \kappa_1)\right).
\end{align}
\end{widetext}

\bibliography{bibliography}